\documentclass[useAMS,usenatbib]{mn2e}

\usepackage{graphicx}
\usepackage{amsmath}
\usepackage{amssymb}

\title[Rapidly varying accretion and AGN feedback]{Rapidly varying
accretion and AGN feedback}

\author[E.C.D. Pope] {Edward
C.D. Pope$^{1}$\thanks{E-mail:e.c.d.pope@leeds.ac.uk}\\ $^{1}$School
of Physics \& Astronomy, University of Leeds, Leeds, UK, LS2 9JT\\}

\begin{document}

\pagerange{\pageref{firstpage}--\pageref{lastpage} \pubyear{2007}}

 \maketitle

\label{firstpage}

\begin{abstract}
Accretion rates onto AGN are likely to be extremely variable on short
timescales; much shorter than the typical cooling time of X-ray
emitting gas in elliptical galaxies and galaxy clusters. Using the
Langevin approach it is shown that, for a simple feedback system, this
can induce variability in the AGN power output that is of much larger
amplitude, and persists for longer timescales, than the initial
fluctuations. An implication of this is that rich galaxy clusters are
expected to show the largest and longest-lived
fluctuations. Stochastic variations in the accretion rate also mean
that the AGN injects energy across a wide range of timescales. This
allows the AGN to maintain a much closer balance with its surroundings
than if it was periodically activated. The possible non-linear
correlation between Bondi accretion rate and jet power, found by
\cite{allen06a}, can be explained if the instantaneous accretion rate,
scaled by jet power, varies log-normally. This explanation also
implies that the duty cycle of AGN activity increases with the
radiative losses of the surroundings, in qualitative agreement with
\cite{best05}.

\end{abstract}


\section{Introduction}

It is well known that elliptical galaxies are commonly the hosts of
powerful radio AGNs \citep[e.g.][]{mclure}. These sources give rise to
lobes of radio emission embedded in the X-ray emitting gaseous haloes
surrounding the host galaxies. There is also growing theoretical
evidence that outflows from AGNs play a significant role in the
evolution of their surroundings \citep[e.g.][]{nature,croton05,
sijacki06, bower06}.

In theoretical work, outflows from AGNs are commonly used to prevent
catastrophic radiative cooling in both elliptical galaxies
\citep[e.g.][]{tabor93,bintab} and galaxy clusters
\citep[e.g.][]{bub01,nature,brueggen03, basson03, omma04,
vecchia04,cat07}. However, for there to be a long term balance between
the heating and cooling processes there must be some sort of feedback
mechanism through which the cooling gas triggers AGN outbursts. The
level of jet activity appears to be a function of the environment and
black hole mass \citep[e.g.][]{burns, best05, best07}. In addition,
observations also suggest that the heating must be relatively gentle
and well matched to its environment \citep[e.g.][]{fab06}.

In numerical simulations, the jet power of an AGN is often related
directly to the effective inflow rate of material through an inner
boundary of the computational grid \citep[e.g.][]{vern05,cat07}. In
reality the picture is not so simple since the overall accretion rate
onto a supermassive black hole, located in an ellitical galaxy, is
probably a combination of its local, and large-scale, environments
\citep[e.g.][]{soker06,hard07, cen07}. The local accretion rate is
probably governed by Bondi accretion
\citep[e.g.][]{allen06a}. However, numerical simulations suggest that
Bondi accretion is not steady, but can be highly variable
\citep[e.g.][and references therein]{edgar04}. We also observe
extremely powerful, cluster-scale outbursts, e.g. Hydra-A
\citep[][]{nulsen05}. These outbursts are probably triggered by
accretion of material from the wider cluster environment, rather than
Bondi accretion. In this scenario, one could envisage cold `blobs' of
material which drop out of the flow around a galaxy, `diffusing'
towards the central black hole. Clearly this process would be expected
to be intrinsically variable.

Most of the infalling material will have appreciable angular momentum;
somehow this gas must be funnelled down onto the black hole,
presumably through an accretion disc. This process will mediate the
accretion rate from the external system that surrounds the black
hole. Depending on the disc processes, the instantaneous black hole
accretion rate need not be strongly correlated with the external
(systemic) accretion rate.

Another source of accretion rate variability would be tidal
interactions and mergers with nearby galaxies
\citep[e.g.][]{hopkins06a}. Furthermore, the dynamic state of the
accreted medium will also affect the black hole accretion rate. In
this respect, some advances in numerical models have been made using
initial conditions based on the simulation of the \cite{s2} by
\cite{heinz06a}.

In short, it is easy to understand how the accretion rate onto an AGN
could be highly variable, on short timescales, compared to the
approach assumed in many hydrodynamic simulations. However, adequately
resolving the required range of scales is clearly impractical with
current computing constraints: most numerical simulations of AGN
feedback do not even resolve scales as small as the Bondi radius, let
alone the accretion disc.

Although there is not extensive literature, the importance of random,
or stochastic, accretion is slowly becoming more
apparent. \cite{hopkins06} employ the phenomenon to investigate the
fueling of low-level AGN, and its effect on the black hole - bulge
relations. In their model, they assume that cold gas stochastically
accretes onto a central supermassive black hole at a rate set by the
dynamics of that gas. In particular, clouds of molecular gas collide
with the supermassive black hole. This provides a distinction between
local, low-luminosity quiescent AGN activity, and violent,
merger-driven bright quasars.

\cite{kp06} and \cite{kp07} also study the effect of small-scale,
randomly oriented accretion event, as a mechanism for feeding nearby
AGN. \cite{kp06} conclude that supermassive black holes can grow
rapidly if most of the mass comes from a sequence of randomly oriented
episodes whose angular momenta are no larger than the angular momentum
of the black hole. This means that the black hole has a low spin, and
therefore a low radiative efficiency, so rapid growth is possible
without exceeding the Eddington limit. Their model requires the
accretion hydrodynamics to be chaotic at the level of the size of the
disc: $\lesssim 0.1$pc. This model is extended by \cite{nayash} who
find that, due to the random orientation of the accretion disc in
these events, energy can be fedback into the surroundings in an almost
isotropic manner. This is important since it is much more efficient
than anisotropic heating.

\cite{nipbin} also argue that the mechanical luminosity of an AGN is
expected to fluctuate across a wide range of timescales, while the
X-ray luminosity of the cluster gas will only vary slowly and weakly
with time. Since energy injection by AGN is expected to balance the
radiative losses of the cluster gas, it makes sense to construct a
dimensionless variable where the AGN luminosity, $H$, is scaled by the
radiative losses, $L_{\rm X}$. This eliminates variations in $H$ which
reflect the size of $L_{\rm X}$. The authors argue that the
probability density function of $\ln(H/L_{\rm X})$, at any time, is
Gaussian; in other words, $H/L_{\rm X}$ is lognormally distributed,
which is an example of a skewed probability distribution.

Throughout physics, stochastic variations are often addressed using
Langevin dynamics \citep[e.g.][]{schuss, stoch}. The Langevin equation
is a stochastic differential equation in which force terms are added
to Newton's second law to approximate the effects of neglected degrees
of freedom. One term represents a frictional force, the other a
random, fluctuating force. In this method, a random variable $f$
represents physical processes that happen rapidly and which may be on
smaller spatial scales than are resolved in the model. The principle
assumptions for this fluctuating force are: $f$ is statistically
independent of the physical variable of interest, $X$. The variations
of $f$ are much more rapid than those in $X$, and the mean of $f$ is
zero. Therefore, $f$ represents a source of `white noise'. This
principle can be applied directly to simulations of AGN feedback in
elliptical galaxies and galaxy clusters.

In this letter, we describe the use of the Langevin equation in
predicting the variability of AGN power output, and its dependence on
environment. A more complex feedback model is used to explain the
possible origin of the non-linear correlation between the Bondi
accretion rate and jet power in a sample of elliptical galaxies
reported in \cite{allen06a}. Finally, we summarise the main findings
of this work.

.

\section{Galaxy and galaxy cluster variability models}

Observations suggest that elliptical galaxies and galaxy clusters must
be relatively close to steady state \citep[e.g.][]{fab06} with the
mass of cold gas and new star formation lower than expected
\citep[][]{edge01}, although see \cite{rafferty06}. In addition the
spectra also indicate the mass dropout rates are much lower than
expected from classical estimates \citep[e.g.][]{voigt04}.

We can construct a very simple, but instructive, phenomenological
model to investigate the effect of AGN feedback. Let us assume that the
accretion rate increases if the radiative losses exceed the energy
injection rate, and vice versa,
\begin{equation}\label{eq:eins}
\frac{{\rm d}\dot{m}}{{\rm d}t} = K_1(L_{\rm X}-H),
\end{equation}
where $H$ is the energy injection rate by the AGN, $L_{\rm X}$ are the
radiative losses of the cluster atmosphere, and $K_1$ is a
constant. If we also assume that $H$ and $\dot{m}$ vary on much
shorter timescales than $L_{\rm X}$, we can treat $L_{\rm X}$ as
approximately constant. In the absence of stochastic accretion we can
write $H=K_2\dot{m}$, where $K_2 = \eta c^{2} \approx 10^{20}{\rm
cm^{2}s^{-2}}$. Solving equation (\ref{eq:eins}), assuming that the
accretion rate is zero at $t=0$, gives,
\begin{equation}\label{eq:zwei}
\dot{m}(t) = \frac{L_{\rm X}}{K_2}\bigg[1 -
\exp\bigg(-\frac{t}{\tau}\bigg)\bigg],
\end{equation}
where $\tau \equiv 1/(K_1 K_2)$. In other words, the system tends
towards a state with a steady accretion rate of $L_{\rm X}/K_2$, as
time passes.

If there is a time-delay in the AGN response we would expect the
accretion rate to overshoot the steady-state value, and subsequently
perform damped sinusoidal oscillations around this point. Of course,
if the time-delay is so large that the overshoot is equal to the
steady-state accretion rate, then the system will be unstable.

The value of $K_1$ is slightly harder to evaluate; we estimate it in
the following way,
\begin{equation}\label{eq:drei}
\frac{{\rm d}\dot{m}}{{\rm d}t} \sim \frac{\dot{m}}{t_{\rm cool}} =
K_1 L_{\rm X}.
\end{equation}
We can write $L_{\rm X} = 5k_{\rm b}T\dot{m}_{\rm c}/(2\mu m_{\rm
p})$, where $\dot{m}_{\rm c}$ is the mass dropout rate from the ICM,
which is distinct from $\dot{m}$ - the accretion rate onto the black
hole. Thus we have
\begin{equation}\label{eq:vier}
\tau \sim 4 \times 10^{3} \bigg(\frac{T}{10^{7}{\rm
K}}\bigg)\bigg(\frac{t_{\rm cool}}{10^{8}{\rm
yrs}}\bigg)\bigg(\frac{\dot{m}_{\rm c}}{\dot{m}}\bigg) \,{\rm yrs}.
\end{equation}
We might reasonably expect $\dot{m}_{\rm c}/\dot{m} \sim 100$
\citep[e.g.][]{pope06} so that $\tau \sim 10^{5}$yrs. This is much
shorter than typical radiative cooling times, thus justifying our
initial approximation that the accretion rate varied on comparatively
short timescales. This also indicates that the heating response
quickly establishes an equilibrium, but that $\tau$ increases with the
temperature of the ICM, suggesting that more massive clusters have a
longer response time.

Now, if we also consider the effect of stochastic accretion, we can
write $H = K_2\dot{m} + \Delta H$, where $\Delta H$ is a random
variable that can be positive or negative. Since the system also has a
steady-state accretion rate, $\langle\dot{m}\rangle = L_{\rm X}/K_2$,
we can write $\dot{m} = \langle\dot{m}\rangle + \Delta \dot{m}$,
\begin{equation}\label{eq:funf}
\frac{{\rm d}\Delta \dot{m}}{{\rm d}t } = -\frac{\Delta \dot{m}}{\tau}
- K_1\Delta H(t).
\end{equation}
Equation (\ref{eq:funf}) has the form of the Langevin equation where
$-K_1\Delta H = \sigma_{\rm N}f(t)$. $f(t)$ is a Gaussian
`white-noise' process, with zero mean and unit standard deviation;
$\sigma_{\rm N}$, is the standard deviation of the noisy process. The
variance of $\Delta \dot{m}$ is given by,
\begin{equation} \label{eq:seben}
\sigma^{2} = \frac{\sigma^{2}_{\rm N}\tau}{2},
\end{equation}
which demonstrates that seed fluctuations of
$\langle\dot{m}\rangle/\sqrt{\tau} \ll \langle\dot{m}\rangle$ can
produce AGN power fluctuations that are comparable with the mean
accretion rate.

The transformation of a random variable through a differential
equation, to obtain a probability distribution, is notoriously complex
\citep[see][for an example]{chat02}. However, in the case described
the probability distribution is Gaussian,
\begin{equation}\label{eq:zehn}
g(\dot{m}){\rm d}\dot{m} = \frac{1}{\sqrt{\pi\sigma^{2}_{\rm
N}\tau}}\exp\bigg[-\frac{(\dot{m}-\langle\dot{m}\rangle)^{2}}{\sigma^{2}_{\rm
N}\tau}\bigg],
\end{equation}
where the width of the distribution grows with $\tau$. Equation
(\ref{eq:vier}) shows that $\tau$ increases with cluster temperature,
which in turn, increases with cluster mass. Therefore, from equation
(\ref{eq:seben}) we should expect to observe greater variability in
more massive systems. However, the ratio
$\langle\dot{m}\rangle/\sigma$ may decrease with increasing cluster
mass, although it is not clear how $t_{\rm cool}$ and
$\dot{m}/\dot{m}_{\rm c}$ scale with cluster mass.

Discretising equation (\ref{eq:funf}) gives the standard form of a
first-order auto-regressive process \citep[e.g.][]{stoch, king03},
\begin{equation} \label{eq:ar2}
X_{\rm t} = \bigg(1-\frac{\Delta t}{\tau}\bigg) X_{\rm t-1} +
\sigma_{\rm N} \Delta t f(t),
\end{equation}
where $\Delta t$ is the time-step of integration. This is a
first-order auto-regressive process because the variable at time $t$
depends on only the value at $t-\Delta t$, and is only physically
meaningful if $ 0 \le 1-\frac{\Delta t}{\tau} \le 1$. Using equation
(\ref{eq:ar2}) we can generate $X_{\rm t}$ as a function of time,
variance given by \citep[e.g.][]{stoch},
\begin{equation}\label{eq:var}
\sigma^{2}_{\rm X} = \frac{\sigma^{2}_{\rm N}}{1-\phi^{2}},
\end{equation}
where $\phi = 1-\Delta t/\tau$. Note that, in this case $\Delta t =
1$, and in the limit that $\tau \gg 1$, $\sigma^{2}_{\rm X} =
\sigma^{2}_{\rm N}\tau/2$, in agreement with equation
(\ref{eq:seben}). Two light curves generated using equation
(\ref{eq:var}) are shown in figure 1.

\begin{figure*}
\begin{minipage}[b]{.4\linewidth}
\centering\includegraphics[width=0.7\linewidth]{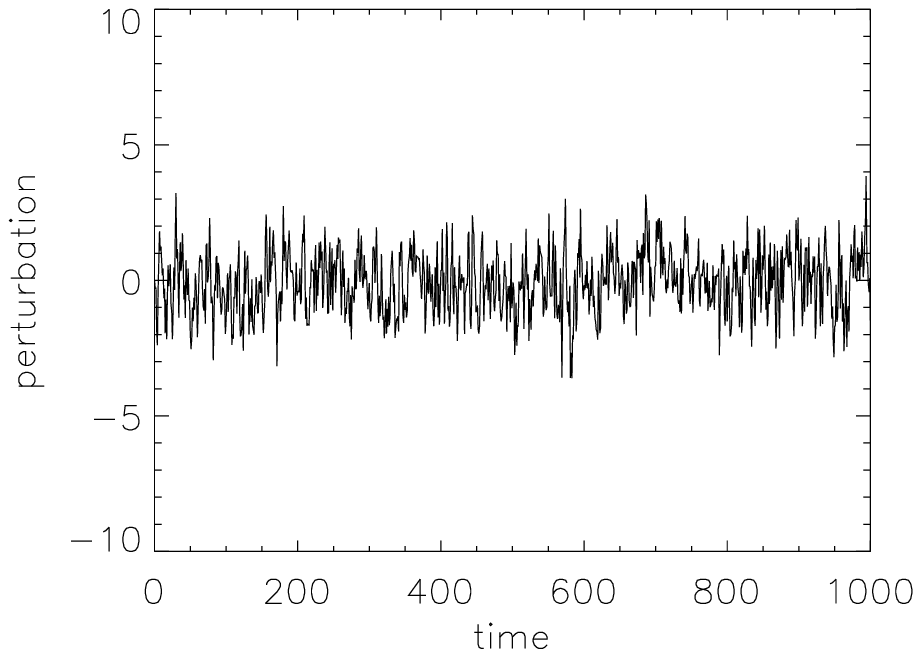}
\end{minipage}\hfill
\begin{minipage}[b]{.4\linewidth}
\centering\includegraphics[width=0.7\linewidth]{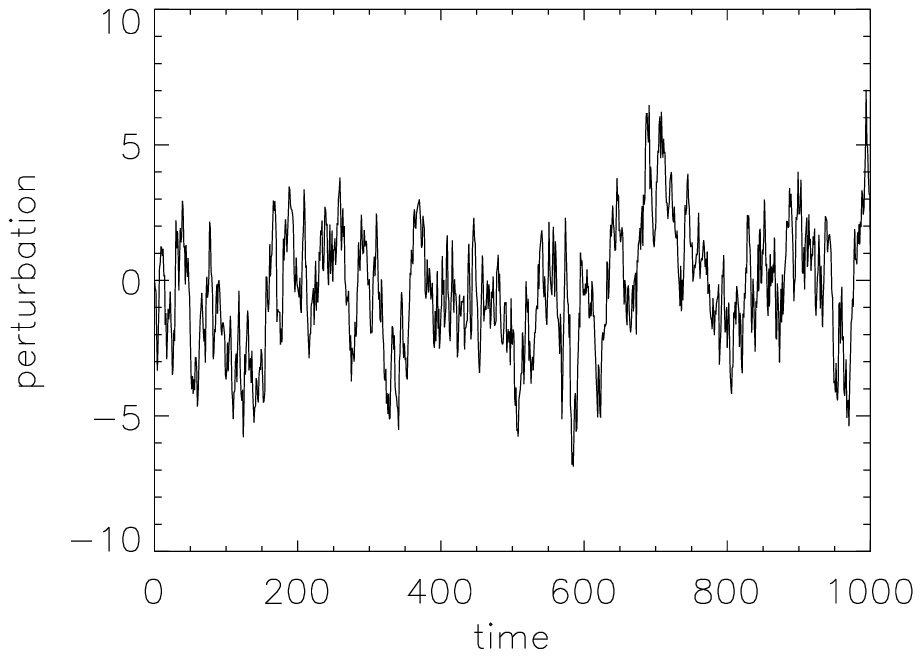}
\end{minipage}\hfill
\caption{Comparison of perturbations ($X_{\rm t}$, in model equations)
generated with the same source of noise, but for different values of
$\phi$, where the standard deviation of the noise is unity. On the
left $\phi = 0.5$, corresponding to $\Delta t/\tau = 1/2$. On the
right $\phi = 0.9$, corresponding to $\Delta t/\tau = 1/10$. Notice
that for larger $\phi$ the range of the variability is greater and
occurs on longer timescales.}
\label{fig:lc}
\end{figure*}

Even if the AGN power output is predominantly described by a Gaussian
probability distribution, the distribution is likely to be truncated
somewhat. This is because the minimum AGN power is zero, while the
maximum steady power output is governed by the Eddington limit. The
truncation of the distribution will be negligible if the width of the
probability distribution is small compared to the mean. However, this
is probably not the case, otherwise we would expect to observe a high
proportion of objects injecting energy at this rate. Thus, the
distribution could only be symmetric if the Eddington limit is equal
to twice the time-averaged AGN power output, which is approximately
equal to $L_{\rm X}$. A simple inspection of the available data
\citep[e.g.][]{smbh04} demonstrates that this is probably not the
case.

\subsection{Non-linear, asymmetric feedback}

In equation (\ref{eq:eins}) we assumed that $L_{\rm X}$ varied slowly,
so that it could be treated as approximately constant. However, if the
radiative losses are not matched by AGN we heating we would reasonably
expect that the radiative losses to increase. The growth of radiative
losses is a positive feedback effect, so it is easy to see that the
system overall could be non-linear and asymmetric for perturbations
about the current, and temporary, steady-state. For example
\citep[e.g.][]{storch},
\begin{equation}\label{eq:non}
\frac{{\rm d}X}{{\rm d}t} = c_{0}-c_{1}X -c_{2}X^{2} + f,
\end{equation}
where $c_{\rm i}>0$, and $f$ is the white-noise random variable.

Regardless of these considerations, given that the accretion rate is
positive, definite, it seems likely that a skewed probability
distribution will result, whatever its origin.

In equation (\ref{eq:non}), the damping is stronger to the lower side
of the equilibrium state, than for deviations to the larger side. As a
result the probability density function (pdf) is no longer Gaussian,
but will be postively skewed and deviations to larger values are more
likely than to smaller. To better understand this, we can think of the
feedback equation as a randomly forced, classical particle moving in a
potential, $U$. The deterministic forcing component is given by the
negative derivative of the potential, $\frac{{\rm d}X}{{\rm d}t} =
-\frac{{\rm d}U}{{\rm d}x} + f$, thus for equation (\ref{eq:funf}): $U
= X^{2}/(2\tau)$ is symmetric, so we would expect the probability
distribution of $X_{\rm i}$ also to be symmetric. In contrast, for
equation (\ref{eq:non}): $U = -c_{0}X + (1/2)c_{1}X^{2} +
(1/3)c_{2}X^{3}$. This is clearly asymmetric and the pdf of the $X{\rm
i}$ values will be skewed towards the less steep slope.

\section{The correlation between instantaneous accretion rate and jet power}

AGN jets are commonly observed to inflate cavities in the X-ray
emitting gas that surrounds galaxies and permeates cluster. The jet
powers can be estimated by dividing the cavity enthalpy by a
characteristic timescale, \citep[e.g.][]{birzan, dunn05, allen06a},
\begin{equation} \label{eq:jets}
P_{\rm jet} = \frac{\gamma}{\gamma-1}\frac{PV}{t},
\end{equation}
where $P$ is the ambient pressure, $V$ is the volume of the cavity,
$\gamma$ is the adiabatic index of the material within the cavity, and
$t$ is the timescale. The pressure, and volume can be estimated from
X-ray observations of galaxies and clusters. Note that this assumes
that the cavity was inflated slowly, and in approximate pressure
equilibrium with its surroundings. 

The appropriate timescale is not clear, and there are several options
to choose from: the time taken to buoyantly rise to the observed
location, the time taken to refill the displaced volume as the cavity
rises, and the sound crossing time for ICM material to traverse a
distance comparable with the bubble radius.

The jet power, as given by equation (\ref{eq:jets}), represents a
time-averaged quantity over the duration of the cavity inflation. As a
result it should correlate well with the mean accretion rate during
this period. \cite{allen06a} calculate the jet powers for a sample of
9 nearby X-ray luminous elliptical galaxies with good optical velocity
dispersion measurements. The optical velocity dispersion measurements
permit black hole mass estimates via the $M_{\rm bh}-\sigma$ relation
\citep[e.g.][]{tremaine}. From this, and the X-ray data,
\cite{allen06a} also estimate the Bondi accretion rate in the vicinity
of each supermassive black hole. It should be stated here that the
value of the Bondi accretion rate is an instantaneous value which
could vary significantly from the mean. They find the following
correlation between the Bondi accretion power and jet power,
\begin{equation} \label{eq:allen}
\log (P_{\rm Bondi}/10^{43}{\rm erg\,s^{-1}}) = 0.65 + 0.77 \log
(P_{\rm jet}/10^{43}{\rm erg\,s^{-1}})
\end{equation}
Interestingly, the correlation is not linear which may simply be a
consequence of the relatively small sample, or a systematic error in
calculating the jet power, for example. Alternatively, the non-linear
correlation may indicate that other important physical effects are
operating: perhaps this is an indication of feedback between the jet
and its fuel supply, or that the jet production efficiency is a
function of jet power \citep[e.g.][]{nemmen}.

The possible explanation presented here is that the non-linear
correlation arises directly from the distribution of the fluctuations
in the instantaneous accretion rate. If the probability distribution
of the accretion rate is symmetric about the mean then we would
expect, on average, to observe a linear correlation between jet-power
and instantaneous accretion rate. As a result this symmetric model of
feedback given in equation (\ref{eq:eins}) cannot provide an
explanation for the possible non-linear correlation between jet power
and Bondi accretion rate reported by \cite{allen06a}, even if the
variance of the distribution is a function of the mean accretion
rate. However, a skewed probability distribution of accretion rates
may be a possible explanation, since the mode and mean will not be
identical. Yet, even this fact is not enough to explain the observed
non-linear correlation: if $P(\dot{m}/\mu \ge 1) = {\rm constant} \ne
0.5$, and not a function of $\mu$ (or the variance) the correlation
between jet power ($\mu$) and instantaneous accretion rate $\dot{m}$
will still be linear. If $P(\dot{m}/\mu \ge 1) > 0.5$, the best-fit
line determined from observations may over-estimate the normalisation
of the relationship since we expect to observe larger accretion rates
more frequently. The reverse is also true. However, if the skewness of
the probability distribution is not constant, but is a function of
$\mu$, this could lead to the observed non-linear correlation.

\cite{nipbin} assumed that the X-ray luminosity of the cluster gas was
slowly varying, while the AGN power varied much more rapidly. In a
similar manner we will assume that the jet power, estimated using
equation (\ref{eq:jets}), is also slowly varying compared to the
instantaneous Bondi accretion rate. Following the same arguments as
\cite{nipbin} further, it makes sense to define the variable $y$ as
the ratio of the accretion power obtained from the instantaneous Bondi
accretion rate, $P_{\rm Bondi}$ and the jet power $P_{\rm jet}$,
\begin{equation}
y = \frac{P_{\rm Bondi}}{P_{\rm jet}}.
\end{equation}
Again, we can argue that $y$ is log-normally distributed,
\begin{equation}
g(y) = \frac{1}{\sqrt{2\pi}S y}\exp\bigg[-\frac{(\ln y -
M)^{2}}{2S^{2}}\bigg],
\end{equation}
for $y > 0$, where $M$ and $S$ are the mean and standard deviation of
$\ln y$. The expectation of $y$ is
\begin{equation}
\langle y \rangle = \exp\bigg(M + \frac{S^{2}}{2}\bigg).
\end{equation}
This quantity can be directly estimated from the observed best-fit
relation between $P_{\rm Bondi}$ and $P_{\rm jet}$. Taking the Bondi
accretion rate-Jet power relation from equation (\ref{eq:allen}) and
assuming that $M=0$, for simplicity, we find,
\begin{equation}
\langle y \rangle \approx {\rm best-fit}\bigg(\frac{P_{\rm
    Bondi}}{P_{\rm jet}}\bigg) = 3.4\times10^{10}P_{\rm jet}^{-0.23} =
    \exp\bigg(\frac{S^{2}}{2}\bigg).
\end{equation}
This tells us that the variance of the distribution is a function of
jet power,
\begin{equation} \label{eq:var}
S^{2} \approx 3 - 0.46 \ln \bigg(\frac{P_{\rm jet}}{10^{43}{\rm
erg\,s^{-1}}}\bigg),
\end{equation}
so that $S^{2} \sim 3$. Such values are comparable with those found by
\cite{nipbin} in their model.

The duty cycle for a source with light curve $L(t)$ is defined by
\cite{cfq} as,
\begin{equation}\label{eq:duty}
\delta \equiv \frac{\langle L \rangle ^{2}}{\langle L^{2} \rangle} =
\exp(S^{2}).
\end{equation}
Substituting equation (\ref{eq:var}) into equation (\ref{eq:duty})
tells us that the duty cycle is also a function of jet power (probably
indirectly),
\begin{equation}\label{eq:duty2}
\delta \approx 0.05\bigg(\frac{P_{\rm jet}}{10^{43}}\bigg)^{0.46}.
\end{equation}
Since there is probably a close balance between the jet power and the
radiative losses of the X-ray halo around the galaxy we may assume
that $\langle P_{\rm jet} \rangle \approx L_{\rm X}$. Therefore, we
can argue that this result strongly suggests that AGN in larger
galaxies, hence larger $L_{\rm X}$, are active for a greater fraction
of the time. In fact, in elliptical galaxies $L_{\rm X}$ is roughly
proportional to the square of the mass of the black hole, $m_{\rm
bh}$, \citep[e.g.][]{best05} at the centre of the galaxy. Therefore we
might expect $\delta \propto m_{\rm bh}^{0.9}$ in elliptical galaxies.

In reality the observed non-linear correlation given in
\cite{allen06a} may not be real, but only a statistical
anomaly. Nevertheless, it is interesting to note that the implications
for the duty cycle qualitatively agree with the observational result
of \cite{best05}. Their result shows that the fraction of AGN in
elliptical galaxies, which are detected in the radio, increases with
the mass of the black hole at the centre of the galaxy. This can be
interpreted probabilistically as the fraction of time the AGN is
active, i.e. the duty cycle. At low black hole masses
($10^{6}-10^{7}{\rm M_{\odot}}$), the observed relationship given by
\cite{best05} scales as $m_{\rm bh}^{1.6}$, although this flattens
sigificantly at higher masses. Interestingly, the estimate of the duty
cycle presented here applies for black hole masses of $\sim 8-9 \times
10^{8}\,M_{\odot}$, which agrees qualititively with the flattening in
the \cite{best05} relation at large black hole masses.

\subsection{Stochastic variability and gentle AGN heating}

Under certain circumstances, it is possible to imagine that AGN
activity is triggered periodically, or exhibits periodic activity of
some sort. \footnote{In an alternative model we could replace $(L_{\rm
X}-H)$ with $\int(L_{\rm X}-H){\rm d}t$ in equation (\ref{eq:eins}),
thus the AGN power output would vary sinusoidally between $0$ and
$K_{2}L_{\rm X}/K_{1}$. The period of the AGN variability would be
$\tau = 2\pi/\sqrt(K_{1}K_{2})$. In this case $K_1$ would have a
different value to the model described in the main text, but would
still probably increase with galaxy, or cluster, mass.} This could
lead to large variations in the temperature, and thus large deviations
from equilibrium. Yet, observations of metal distributions and cooling
time profiles in clusters seem to suggest that there is an extremely
close balance between heating and cooling in galaxy clusters. This is
what one would expect from continuous feedback-controlled
heating. However, in the simplest case the ensuing steady-state could
result in the AGN being permanently active with a constant power
output \citep[e.g.][]{hoeft}. (This example is of interest considering
the temporal form of the accretion rate, which appears very similar to
equation (\ref{eq:eins}), in some cases). However, this is not what we
expect from the stochastic feedback model discussed above. Instead,
the rapid stochastic variations in the accretion rate mean that the
AGN injects energy across a wide range of timescales. Thus, the AGN
can maintain a much closer balance with its surroundings than if it
was periodically activated.

For a skewed probability distribution of AGN power output, much of
this heating must occur at relatively low powers and so will provide
the gentle heating that is consistent with observations. The power
spectral density of the equation (\ref{eq:funf}) described above is
typical of a `red-noise' process \citep[e.g.][]{vaughan}, with
$\Gamma(\omega) \approx \frac{a\sigma^{2}_{\rm N}}{b + \omega^{2}}$,
where $a$ and $b$ are constants, and $\omega$ is the angular frequency
of the variations.

\subsection{Stability of the system}

Time-delay in control systems tends to reduce the stability of the
system. Indeed, for sufficiently long delays the system will be
completely unstable. In the galaxy cluster example, if the heating
response from the AGN is longer than the local cooling time, the
system is unlikely to be stable.

The main components that contribute to the time-delay are: 1) the time
taken for inflowing gas to reach the black hole, and 2) the time taken
for the energy fedback into the ICM to be dissipated. Both components
are difficult to estimate, but the former moreso since it is probably
composed of at least two stages: the inflow from large radii, and
inflow through an accretion disc. For example, the gas presumeably has
some angular momentum, and should stagnate at a radius
\citep[e.g.][and references therein]{sarazin},
\begin{equation}
r_{\rm st} = \frac{l^{2}}{GM(<r_{\rm st})},
\end{equation}
where $l$ is the angular momentum per unit mass of the gas, $G$ is
Newton's gravitational constant, and $M(<r_{\rm st})$ is the mass
within the stagnation radius. This radius is expected to be of the
order of a few kiloparsecs \citep[][]{sarazin}. Dynamical interactions
could then scatter the gas clouds and allow a fraction to move closer
to the black hole.

The majority of the material will pass through an accretion disc that
encircles the black hole; the inflow timescale being given by the
viscous drift timescale $t_{\rm vis} \sim r^{2}/\nu$, where $\nu$ is
the viscosity and $r$ is the radial distance from the black
hole. \cite{czerny} gives this time as $\sim 1$Myr at roughly 0.1 pc
for a disc around a $10^{8} M_{\odot}$ black hole.

The time taken to dissipate the injected energy is also unknown, but
it is concievable that the heating response of an AGN to a particlar
event can have a significant time lag which could destabilise the
system. This could have dramatic consequences were it not for the
presence of small stochastic fluctuations which can give rise to
extremely powerful, long lived AGN outbursts. Such outbursts can occur
at any time and may be sufficient to stabilise the system. There is a
slight irony in this possibility since in other systems, stochastic
variability is usually thought of as a destabilising force.

\section{Summary}

There are many processes in the accretion of material on to an AGN
that occur on relatively short timescales compared to the cooling time
of the X-ray emitting gas in and around elliptical galaxies and galaxy
clusters. In a feedback system these stochastic variations have been
shown to lead to much longer and larger fluctuations in the AGN power
output. Larger fluctuations occur in systems with longer response
times, e.g. rich galaxy clusters. The rapid variability also means
that power is injected across a range of timescales allowing an
intimate balance between heating and cooling.

The possible non-linear correlation between Bondi accretion rate and
jet power in the \cite{allen06a} sample can be explained by a skewed
probability distribution for the ratio of the instantaneous accretion
rate to the jet power. This explanation also predicts that the
duty-cycle of AGN activity varies with its environment, or black hole
mass.

Stochastic variability of the accretion rate may help to stabilise the
system against the large time-delays that are probably inherent with
AGN feedback.

\section{Acknowledgements}

The author would like to thank Christian Kaiser, David Pope, Georgi
Pavlovski, Jim Hinton and Willem-Jan De Wit for useful discussions,
and the anonymous referee for constructive comments.

\bibliography{database} \bibliographystyle{mn2e}

\label{lastpage}

\end{document}